\documentstyle[preprint,aps,12pt]{revtex}
\begin{document}

\input epsf
\draft

\title{LAPW frozen-phonon calculation, shell model
lattice dynamics and specific-heat measurement of SnO}

\author{S. Koval$^a$ \thanks{Present address: International Centre for 
        Theoretical Physics, Strada Costiera 11, 34014, Trieste, Italy.}, 
        R. Burriel$^b$, M.G. Stachiotti$^a$, M. Castro$^b$
        R.L. Migoni$^a$, M.S. Moreno$^d$ \thanks{Present address: Goldaracena 
        740, 2820 Gualeguaychu (Entre Rios), Argentina}, A. Varela$^d$ and 
        C.O. Rodriguez$^e$ }

\address{$^a$ Instituto de F\'{\i}sica Rosario, Universidad Nacional de
              Rosario,\\
              27 de Febrero 210 Bis, 2000 Rosario, Argentina. \\
         $^b$ Instituto de Ciencia de Materiales de Arag\'on, CSIC, Universidad
              de Zaragoza,\\Plaza de San Francisco, 50009 Zaragoza, Spain.\\
         $^c$ Departamento de F\'{\i}sica, TENAES, Universidad Nacional de 
              La Plata, \\ C.C. 67, 1900 La Plata, Argentina.\\
         $^d$ Dpto. de Qu\'{\i}mica Inorg\'anica, Facultad de Ciencias
              Qu\'{\i}micas, Universidad Complutense,\\28040 Madrid, Spain.\\
         $^e$ IFLYSIB, Grupo de F\'\i sica del S\'olido, C.C.565, 
              1900 La Plata, Argentina.} 

\maketitle

\begin{abstract}
An ab-initio Linear Augmented Plane-Wave (LAPW) calculation of 
the zone-centered phonon frequencies of SnO has been performed. 
E$_g$ symmetry has been ascribed to the mode observed at 113 cm$^{-1}$ 
in Raman measurements, discarding a previous B$_{1g}$ assignement.
The other phonon modes measured by Raman spectroscopy are also 
well reproduced. A new shell-model has also been developed,
that gives good agreement of the zone-centered frequencies compared 
to the measured data and the LAPW results. Specific heat measurements 
have been performed between 5 K and 110 K. Computation of the specific 
heat and the M\"{o}ssbauer recoilless fraction with the improved 
shell-model shows a good agreement with the experimental data as 
a function of temperature. \\ 


\end{abstract}

\newpage

SnO has not been subject of extensive investigations due to its
thermal decomposition at relatively low temperatures, precluding 
applications. Presently it is object of a renewed interest, due to
its ability to be an excellent anode material \cite{Ido97} and, so far,
there is no confident description of its vibrational structure. Very 
recently, three contributions were concerned with an appropiate 
description of its electronic properties \cite{Lef98,Mey98,Pel93}. However, 
a proper description of its phonons and lattice dynamics is absent.

A shell model for the lattice dynamics of SnO has been developed recently with
the aim to analyse the M\"{o}ssbauer recoilless fraction of Sn \cite{sno96}. 
The resulting zone center phonon frequencies have been compared with ab-initio 
full potential linear-muffin-tin-orbital (LMTO) calculations \cite{Pel93} and 
Raman and infrared reflectivity measurements \cite{Geu84}. The agreement was 
in general good for the M\"{o}ssbauer data as well as for the phonon 
frequencies. However, the latter showed some discrepancies. On one hand, the 
analysis of the experimental data assigned the B$_{1g}$ symmetry to a Raman 
peak at 113 $cm^{-1}$, while both the LMTO and shell model calculations gave 
a frequency of $\approx$ 370 cm$^{-1}$ for the B$_{1g}$ mode. Since this mode 
involves only oxygen displacements, it is hard to believe that it has such a 
low frequency as experimentally assigned, because there are other Sn-modes 
with higher frequencies. This inconsistency in the assignment of the 
113 $cm^{-1}$ Raman peak remained unnoticed even in the most recent works 
\cite{San98}. On the other hand, the lowest frequency mode is of E$_g$ 
symmetry, involving mainly Sn displacements, and has a frequency of 160 or 
143 $cm^{-1}$ according to the shell model \cite{sno96} or the LMTO 
\cite{Pel93} calculation, respectively. An additional discrepancy appears for 
the A$_{2u}$ mode, which has a frequency of 255 $cm^{-1}$ for the shell model, 
while the LMTO calculation leads to a value of 396 $cm^{-1}$, and no 
experimental data is available for this mode. 

In order to elucidate these questions we undertake in
this work a full potential LAPW calculation of the zone-center modes, which is 
more confident than the LMTO method for the calculation of phonon energies,
particularly in the quite open structure of SnO. We then refine the shell 
model parameters by fitting them to the LAPW results and the spectroscopic
experimental data. In addition, we present measurements of the specific heat, 
which will allow to check the consistency of the determined 
shell-model parameters. We remark that the development of a reliable model of 
SnO should be interesting for the study of the tweed microstructure associated 
with this intrisically non-stoichiometric compound \cite{defsno}.

The ab-initio calculations were performed within the Local Density 
Approximation (LDA) to density functional theory, using the full-potential
LAPW method. In this method no shape approximation on either the
potential or the electronic charge density is made. We use the WIEN95
implementation of the method \cite{wien} which allows the inclusion
of local orbitals (LO) in the basis, making possible a consistent treatment
of semicore and valence states in one energy window hence insuring proper 
orthogonality \cite{lo}. The Ceperley-Alder parametrization for the 
exchange-correlation potential is used \cite{ca}.

The atomic sphere radii ($R_i$) 2.15 and 1.60 a.u. were used for Sn and O, 
respectively. For the parameter $RK_{max}$, which controlls the size of the 
basis sets in these calculations, we take the value of 8. This 
gives well converged basis sets consisting of approximately 
990 LAPW functions plus local orbitals. We introduce LO to include the 
following orbitals in the basis set: Sn 4$p$ and 4$d$, and O 2$s$.

Integrations in reciprocal space were performed using the tetrahedron-method. 
We used 7$\times$7$\times$5 meshes which represent 30 $k$-points in the 
irreducible Brillouin zone.

We determined the phonon frequencies and eigenvectors of particular symmetry
using the frozen phonon approach, by calculating atomic forces for several 
small displacements ($u \approx 0.3\%$ of the lattice constant) consistent with
the symmetry and small enough to be in the linear regime. From the forces as a
function of displacements the dynamical matrix was constructed and 
diagonalized.

We observe in Table I that in general the LAPW-frequencies are somewhat 
smaller than the LMTO values. The LAPW result for the E$_u$ mode is in better 
agreement with the experiment than the LMTO result. The lowest mode is 
confirmed to be of E$_g$ symmetry as in the LMTO calculation \cite{Pel93},
although it lies considerably lower in frequency. The present LAPW calculation
leads to a frequency for this mode in exact agreement with the position of the
peak observed in the Raman experiment \cite{Geu84}. However the experimental 
analysis leads to an identification of this peak as arising from the B$_{1g}$ 
mode. We believe that this assignement is wrong on the basis of several 
grounds. First, both ab-initio calculations already mentioned predict the E$_g$
symmetry for the lowest frequency mode. Second, the B$_{1g}$ mode involves only
antiparallel z-displacements of O ions, while the A$_{1g}$ mode is the 
analogous for the Sn ions. Both modes are determined by Sn-O interactions, 
while the Sn-Sn and O-O forces play an additional role in the A$_{1g}$ and the 
B$_{1g}$ modes, respectively. Therefore, as long as the Sn-O interaction is 
dominant (as it comes out from the LAPW calculations and shell model results 
which will be described latter), the higher frequency mode will be the one 
corresponding to movements of lighter atoms, i.e. the B$_{1g}$ mode. Third, for
the isostructural compound $PbO$, the measured frequency of the B$_{1g}$ mode
~\cite{Ada92} is close to the calculated value for tin monoxide. 

The frequency of the A$_{2u}$ mode could not be confidently determined. As
a consequence of the general underestimation of the electronic band gap 
within the LDA, we obtain a very small band gap for SnO. When
the atoms are displaced according to the A$_{2u}$ eigenvector pattern, part of
the conduction band crosses below the top of the valence band and thus becomes 
populated. This affects the displacement dependence of the energy, which leads
to an unreliable frequency value of this phonon. The same problem arises from 
the LMTO calculation, which is based also on the LDA, so that the previously
published value is also unreliable.\cite{Pel93}

Now we improve the shell model of Ref. \cite{sno96}. It contains short range 
shell-shell interactions of the Buckingham form for the nearest neighbor pairs
O-O and Sn-O and up to second neighbors for the Sn-Sn pair, as well as long 
range Coulomb interactions among all ions. Starting from the parameter values 
given in Ref. \cite{sno96} we further fitted them to the new LAPW results, in 
addition to the experimental phonon data. The most significant differences, 
which appeared for the E$_g$ modes, could be corrected by modifying the O-O 
interaction potential and the O polarizability. We also adjusted slightly the 
Sn-O interaction to improve the fitting. With the parameter set shown in Table
II we obtain the zone-center phonon frequencies shown in Table I. A 
good agreement with the LAPW results is achieved. Also the 
experimental data are well reproduced, except for 
the 113 $cm^{-1}$ experimental phonon data, as previously discussed. 
Considering the experimental values for the E$_u$ and A$_{1g}$ modes and the
LAPW values for the remaining zone center modes we obtain a RMS frequency
difference of 7 cm$^{-1}$  with our shell model result. The shell
model phonons are stable throughout the Brillouin zone. 

In order to validate the present shell model with experimental data we use it 
to compute both the M\"{o}ssbauer recoiless fraction and the phonon density of 
states (PDOS) which allows a direct calculation of the specific heat. The 
M\"ossbauer recoilless fraction $f(T)$ calculated  with the present model 
shows no significant difference with the previous calculation \cite{sno96}. 

The specific heat of SnO has been measured in a commercial 
adiabatic calorimeter from Termis Ltd. \cite{Pav94} between 5 K and 110 K. An 
amount of 2 g of a finely grounded sample was sealed in the 1 cc sample 
container with 50 mbar of He gas to improve the heat exchange and 
temperature equilibrium. Below 5 K the helium gas was absorbed on the 
sample surface preventing to reach thermal equilibrium within measurable 
times. Carbon-glass and Rh-Fe thermometers were used on the sample 
holder and on the adiabatic shield that surrounds it. The temperature of the 
shield is controlled to follow the sample temperature providing adiabatic 
conditions. The sample specific heat was obtained after substracting the 
empty sample holder contribution measured on a previous experiment. The 
experimental values, shown in Fig. 1a, present a characteristic maximum at 
10 K in the $C_p/T^3$ vs. $T$ representation. The effective Debye 
temperatures, derived from the calorimetric data, have been represented in 
Fig. 1b.

From the calculated PDOS we derived the molar heat capacity using 
the following expresion:
\begin{equation}
C_v(T) = 3nR \int_{0}^{\omega_{max}} g(\omega) 
         E \left(\frac{\hbar \omega}{k_B T} \right) d\omega
\end{equation}
where n is the number of atoms by formula unit, R is the molar gas constant, 
$g(\omega)$ is the normalized PDOS determined by the shell model, and
$E(x)$ is the Einstein function:
\begin{equation}
E(x) = \left(\frac{x/2}{sinh(x/2)} \right)^2
\end{equation}
The low-frequency limit of $g(\omega)$ has been smoothed by fitting a 
quadratic law to the histogram, in order to
avoid the influence of noise in the evaluation of the heat capacity.
To take into account an estimated presence of 10 \% Sn vacancies in the
employed samples, we use a value n = 1.9 for the evaluation of C$_v$ according
to Eq. (1) as well as for the conversion of the experimental data to molar 
units. Note that $C_v \approx C_p$ for a solid, thus we shall compare directly
the calculated $C_v$ with the measured $C_p$.
The calculated specific heat, shown in Fig. 1a, has the maximum shifted
to a higher temperature, 16 K, with a height slightly lower than for the 
experimental data. A similar behaviour can be seen on the efective Debye 
temperatures, represented in Fig 1b, with the minimum slightly shifted with 
respect to the experimental one. The calculations have a remarkable 
coincidence with the experimental measurements above 15 K. The whole 
result is quite good considering that the model corresponds to a perfect 
solid. Cation vacancies and the elastic relaxation  existing in the real 
solid \cite{defsno} have not been taken into account in the model. Also the 
uncertainty in the acoustic and low-frequency modes, which give the most 
important contribution to the specific heat at low temperature, can account 
for the difference between calculated and experimental data.

As general features in our calculation, we observe a small Debye region of 
$\approx$ 4 K (defined as the region where
$\Theta(T) \approx \Theta(T=0$K$)$) and a large drop ($\approx 35\%$ of the 
zero-temperature value) in $\Theta(T)$ at low temperatures. The origin of 
these behaviours may be ascribed to a strong hybridization of the acoustic 
and low frequency optical bands near the center of the Brillouin zone (BZ), 
which can be observed particularly in the (k,k,0) symmetry direction (the 
same result is found in our previous shell model calculation \cite{sno96}). 
This hybridization causes on one hand the onset of curvature
in the acoustic dispersion curves at quite low frequencies
which  accounts for the small Debye region mentioned.
On the other hand, the mentioned hybridization causes
low lying modes towards the end of BZ which are responsible
for the large drop in the Debye temperature. 

Recently, the partial Sn-PDOS has been obtained from measurements 
performed in the Advanced Photon Source at the Argonne National 
Laboratory~\cite{hu98}. The accord with our results is very good in the 
positions as well as in the height of the peaks. Moreover, this accord, 
together with the fact that our calculated
partial O-PDOS has nearly vanishing spectral weight for frequencies
$\omega < 220$ cm$^{-1}$, reinforces our conclusion that the oxygen
mode of B$_{1g}$ symmetry was wrongly assigned in Raman experiments
\cite{Geu84}. These results will be 
published elsewhere together with data on other Sn compounds. \\

Acknowledgements:
Partial support from CICYT Project no. MAT97-0987 is acknowledged.
Support through fellowships and a grant from the Consejo Nacional de 
Investigaciones Cient\'{\i}ficas y T\'ecnicas de la Rep\'ublica Argentina is 
also acknowledged.

\figure{}

\vskip1.3cm
\hskip 0.cm

\begin{figure}[ht]

\begin{center}

\hskip 0.cm
\epsfysize=15cm
\epsfxsize=11cm
\
\epsffile{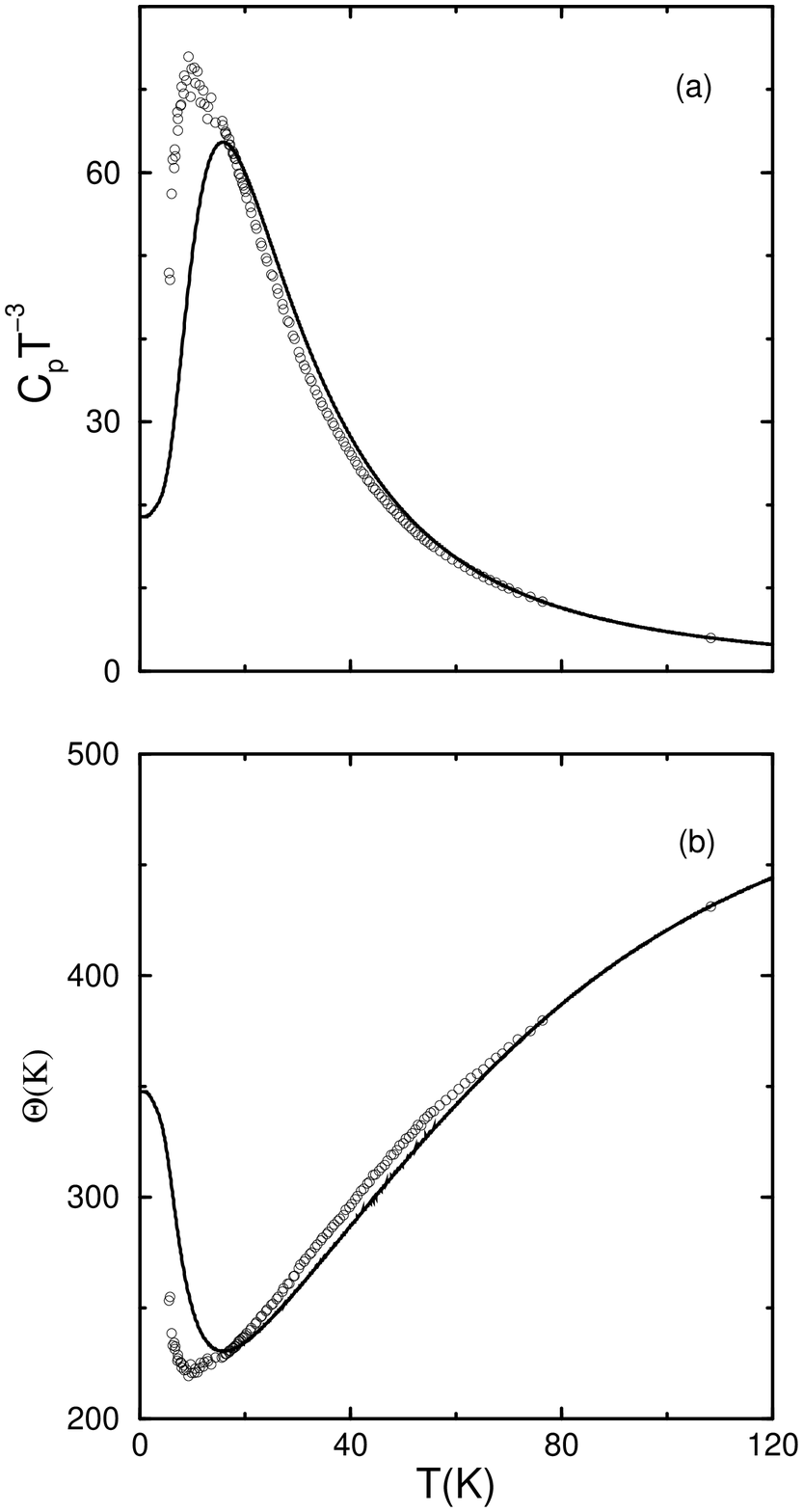}

\vskip 1.cm
\caption{Specific heat (a) and effective Debye temperature (b)
as functions of temperature. Solid line: shell model calculation. Open 
circles: experimental data. The $C_p/T^3$ values are given in units of
$10^{-7} 5.7 R/K^3$}
\label{fig1}
\end{center}
\end{figure}

\newpage

\begin{table}
\caption{Frequencies of the TO zone-center modes of 
$SnO$: for the first-principle LAPW calculation, for the shell model, 
for the LMTO calculation and for the experiment (Unit: $cm^{-1}$). We also show
the relative atomic displacements for each mode, according to the LAPW
calculation. A and B symmetries correspond to c-polarized modes, E symmetries 
to a-polarization. $+(-)$ means in-phase (out of phase) ionic motions.}
\label{table1}

\vskip +.2cm

\begin{tabular}{| c | c | c | c | c | c | c | c | c | }  
 Mode   & LAPW & Shell Model & LMTO~\cite{Pel93}&Experiment~\cite{Geu84} 
                                                 &$Sn+Sn$&$O+O$&$Sn-Sn$&$O-O$\\
\hline
$ E_g$  &   460 &  450    &  494       &          &   0  &  0   & 0.023 &  1  \\
$A_{2u}$&       &  240    &  396 ?     &          &   0  &  1   &   0   &  0  \\
$B_{1g}$&   350 &  357    &  370       &   113 ?  &   0  &  0   &   0   &  1  \\
$E_u$   &   271 &  267    &  296       &   260    &   0  &  1   &   0   &  0  \\
$A_{1g}$&   212 &  211    &  211       &   211    &   0  &  0   &   1   &  0  \\
$E_g$   &   113 &  122    &  143       &          &   0  &  0   &   1   & 0.17\\
\end{tabular}
\end{table}

\newpage

\begin{table}
\caption{Parameters of the model: $a,b,c$: potential parameters;
$Z,Y$: ionic and shell charges; $K$: on site core-shell force constant;
$A,B$: longitudinal and transversal force constants
between neighboring ions.}
\label{table2}

\vskip +.2cm

\begin{tabular}{ c c c c c c  c c c c  } 
Interaction & $a$ (ev) & $b$ (\AA$^{-1})$ & $c$ (eV\AA$^6$) &
                  $A (e^2/v_a)$ & $B(e^2/v_a)$ &
                                           Ion&$Z(|e|)$&$Y(|e|)$&$K(e^2/v_a)$\\
\hline
 Sn-O   &   528 & 2.85  &  0   & 37.05  & -5.9  & O  & -1.4 & -3.5 & 300.52 \\
 Sn-Sn  &  2309 & 2.50  &  0   & 10.0  & -1.1  & Sn &  1.4 &  0   & $\infty$\\
 O-O    & 76031 & 3.925 & 2000 & -0.89 &  7.27 &    &      &      &        \\  
\end{tabular}
\end{table}

\end{document}